# Spectral behaviour and Flaring Activity of II Peg in the Ultraviolet

M.R. Sanad

Astronomy Department, National Research Institute of Astronomy and Geophysics (NRIAG), Helwan - Cairo – Egypt, mrsanad1@yahoo.com

## Abstract

We have investigated the spectral behaviour of II Peg binary system in the ultraviolet band by using International Ultraviolet Explorer (IUE) observations over the period (1979– 1993).The ultraviolet observations show indication of flare activity in the chromosphere and transition region with their enhanced spectral lines. Before and after the flare activity the ultraviolet spectral lines show low, intermediate and high flux. The spectral behaviour is compared with previous studies. We detect prominent flare activity in (1989, 1990, 1992). Before and after this period there is a gradual clear decrease in the level of activity. The reddening of II Peg was estimated from 2200 Å absorption feature to be E (B-V) = 0.10 ± 0.02.We determined the mean rate of mass loss to be ~ $1 \times 10^{-8}$ $M_\odot$ yr$^{-1}$, and the ultraviolet luminosity to be ~ $6 \times 10^{29}$ erg s$^{-1}$.We attributed the spectral variations to a cyclic behaviour of the underlying magnetic dynamo and the prominent activity can be interpreted by the model of two – ribbon flare.

## 1. Introduction

RS Canum Venaticorm Variable Stars (RS CVn) are active chromospheric binary systems with large starspots, intense chromospheric plages, coronal X-ray emission and microwave emissions, as well as enhanced flares in the all wavelengths from optical, radio, ultraviolet and X - Ray regions (Sanad et al. 2021; Hall 1981).

II Peg is an active rotating single line spectroscopic binary system with an orbital period 6.7 days (Ajaz Ahmad Dar et al. 2018). II Peg was distinguished as BY Dra variable system but a study of Rucinski (1977) indicated that it was an RS Canum Venaticorum Variable (RS CVn). The binary system consists of (II Peg A) a cool subgiant orange K type star and has started to expand and evolve from the main sequence (Berdyugina et al. 1998). Starspots cover nearly 40 % of the surface of II Peg A and produces strong flares observed at all wavelengths (Covino et al. 2000, Schwartz 1981)and in 2005 a super flare was observed by a Swift Gamma Ray Burst mission, it was a biggest observed stellar flare and more than hundred million times the solar flare *(*Osten, Rachel et al. 2007*)*and its companion (II Peg B) is very close to be observed directly, it is a red dwarf M type main sequence star (Berdyugina et al. 1998).

Rodono et al. (1986, 1987) presented both optical and ultraviolet spectroscopy taken in 1981. The UV observations revealed a variation with phase. This was explained as the rotation of an active region near the surface of II Peg and correlated with the photospheric spots. Byrne et al. (1987) analysed ultraviolet spectra and showed that the conditions in the active region were attributed to a complex of activated magnetic loops.

Doyle et al. (1989) presented ultraviolet spectroscopic observations of II Peg in 1983 and showed evidence for flare activity. They isolated the flare radiation and derived electron pressures from intersystem line flux ratios and estimated the total power of flare output over the whole atmosphere as well as they derived the flaring volumes and estimated a surface filling factor for flares.

In this paper we discuss the spectral behaviour of II Peg binary system by using ultraviolet data obtained from the International Ultraviolet Explorer (IUE). The important observational characteristics in our investigation are that the fluxes of emission lines in the ultraviolet have approximately the similar spectral behaviour indicating a common origination source in the transition region and chromosphere of II Peg system and the activity is confirmed with ultraviolet data, visible and X – ray data.

We discuss the ultraviolet observations in section 2, section 3 reveals the results and discussions of the procedure of estimating the reddening, the spectral behaviour of line fluxes in the emitting source and the calculations of Ultraviolet luminosity and the rate of mass loss, and section 4 shows the conclusions of the investigation.

## 2. IUE Observations and Reduction of Data

The International Ultraviolet Explorer spectra with low resolution have been obtained from the MAST IUE system through its principle centre at https://archive.stsci.edu/iue/. A detailed description of the ultraviolet low-resolution data is found in Rodriguez-Pascual et al. (1999) and Gonzalez et al. (2001).

The observational data were processed using the standard IUEDAC IDL software for the processing of spectra. We referenced the spectra to the orbital phase of the II Peg system using the ephemeris of Ajaz Ahmad Dar et al. (2018).

$$\text{HJD} = 2443030.239 + 6^{d}.724183 \times \text{E} \tag{1}$$

Table 1 lists the ultraviolet observations for II Peg with low resolution. The spectral data were inspected carefully in the 1150 – 1950 Å region to recognise and reject underexposed or overexposed data. The observations of IUE on II Peg covering nearly all orbital phases, it is known that to calculate the phase we need both the starting point of the cycle and the period by applying the known relation:

$$\text{Phase} = \text{decimal part of} \left( \frac{t - t_0}{P} \right) \tag{2}$$

In our study the starting point is the first observation of IUE data (SWP06362) representing zero phase and the period as taken from the reference (Ajaz 2018) and as well known any two phases which differ by integer are the same. The positions of the binary components II Peg A and II Peg B at different phases say 0 or 0.5 will lead to and depend on the values of their fluxes in the beginning of the cycle and half cycle as the orbital motion along the line of sight alternates toward and away i.e. at the beginning of the cycle the flux is high and at half cycle the flux is low and so on with repeated pattern representing periodic variations for all phases.

Some examples of spectral lines are given in Fig. 1, revealing the changes of line fluxes at different times and the flaring activity. The emission lines are originated in the chromospheres and transition region of II Peg A star. Spectral lines with different states of ionization up to NV 1240 Å, Si IV 1400 Å, and CIV 1550 Å have been found in II Peg, the emission is suitable at all phases.

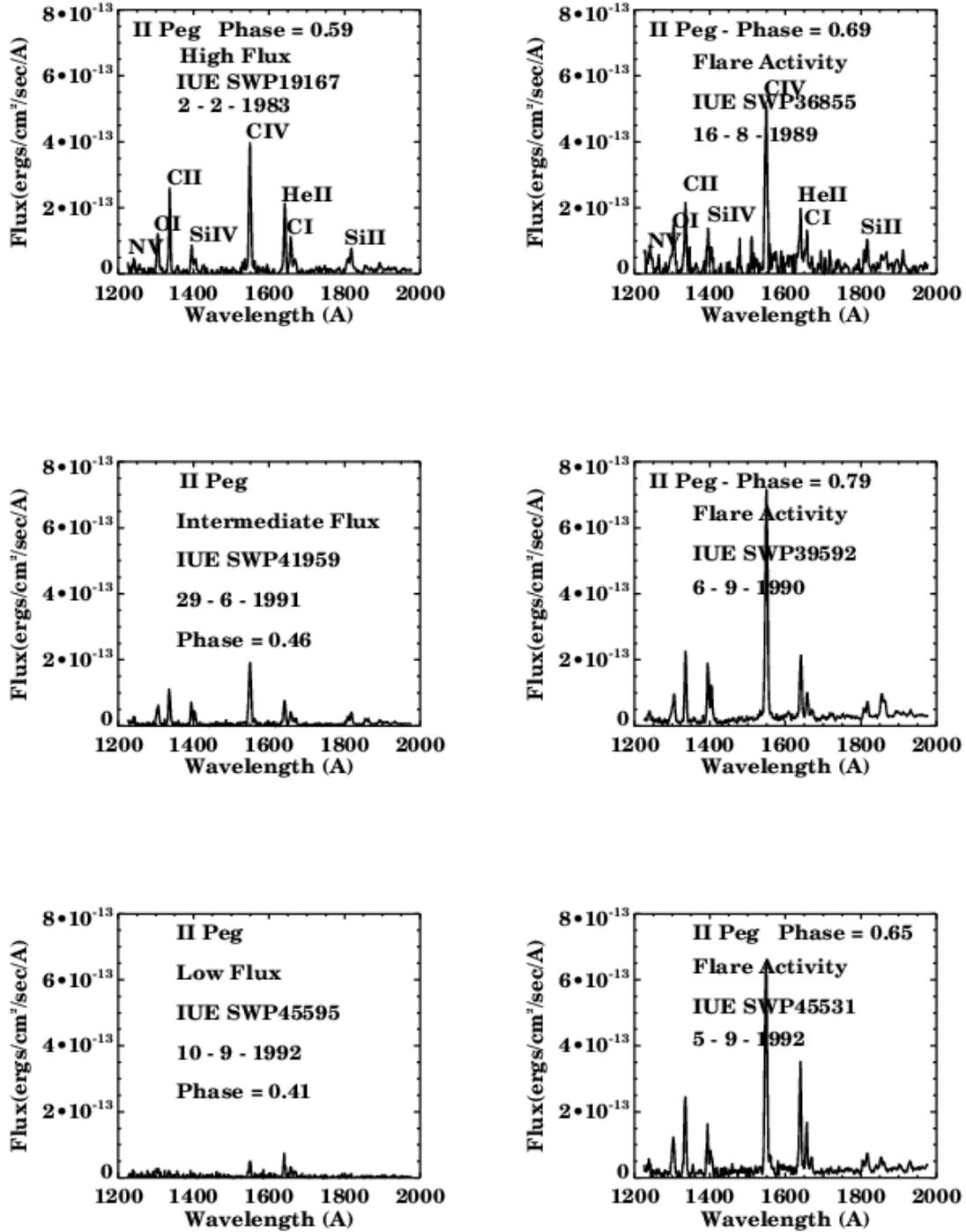

Fig. 1 IUE spectrum of II Peg with high, intermediate and low flux at phases 0.59, 0.46 & 0.41 at the left hand side and the flaring activity at the right hand side

Table 1. Journal of ultraviolet observations of II Peg

| Data ID | Dispersion | Aperture | H.J.D. | Exposure time (s) | Phase |
|---|---|---|---|---|---|
| SWP06362 | Low | Large | 2444118.53512 | 2400 | 0.84 |
| SWP10328 | Low | Large | 2444522.52134 | 6000 | 0.92 |
| SWP15147 | Low | Large | 2444878.51254 | 6000 | 0.86 |
| SWP15166 | Low | Large | 2444880.58741 | 6000 | 0.16 |
| SWP19167 | Low | Large | 2445367.51236 | 3000 | 0.59 |
| SWP19205 | Low | Large | 2445371.25508 | 4800 | 0.14 |
| SWP26457 | Low | Large | 2446268.52541 | 2400 | 0.58 |
| SWP26486 | Low | Large | 2446272.53265 | 3000 | 0.17 |
| SWP29186 | Low | Large | 2446683.52541 | 10800 | 0.30 |
| SWP29215 | Low | Large | 2446687.52543 | 10800 | 0.89 |
| SWP29278 | Low | Large | 2446694.56547 | 2700 | 0.93 |
| SWP36852 | Low | Large | 2447753.52568 | 3600 | 0.42 |
| SWP36855 | Low | Large | 2447754.53654 | 1800 | 0.57 |
| SWP36864 | Low | Large | 2447755.32206 | 3600 | 0.69 |
| SWP39579 | Low | Large | 2448139.29375 | 3600 | 0.80 |
| SWP39584 | Low | Large | 2448140.16816 | 3600 | 0.93 |
| SWP39591 | Low | Large | 2448141.16765 | 4800 | 0.98 |
| SWP39592 | Low | Large | 2448141.34873 | 4800 | 0.79 |
| SWP41959 | Low | Large | 2448466.52541 | 3120 | 0.46 |
| SWP45531 | Low | Large | 2448871.26539 | 1200 | 0.65 |
| SWP45553 | Low | Large | 2448872.56521 | 6000 | 0.84 |
| SWP45573 | Low | Large | 2448874.39836 | 4440 | 0.12 |
| SWP45595 | Low | Large | 2448876.32451 | 1800 | 0.41 |
| SWP45618 | Low | Large | 2448877.51654 | 3600 | 0.58 |
| SWP45661 | Low | Large | 2448881.52541 | 3300 | 0.18 |
| SWP48999 | Low | Large | 2449284.53541 | 1500 | 0.11 |
| SWP49027 | Low | Large | 2449288.51547 | 1500 | 0.70 |
| SWP49222 | Low | Large | 2449305.55413 | 1500 | 0.23 |
| SWP49261 | Low | Large | 2449310.52547 | 1500 | 0.98 |

## 3. Results and Discussions

### 3.1 Reddening determination of II peg

The calculation of reddening of II Peg depends on using the suitable data set for Short Wavelength spectra with resolution (6 Å) in the wavelengths between 1150 - 1950 Å and Long Wavelength spectra with the same resolution in the wavelengths between 2000-3000 Å. The Short Wavelength spectra are binned in 15 Å bins and 25 Å bins for Long Wavelength data. Both data sets give the spectrum shape of a reddened star, with its distinguished depression at 2200 Å. The following observations were selected for our determination of the reddening (SWP06263 – LWR05058) (SWP10328 – LWR08991) (SWP15151 – LWR09002) leading to the best spectrum suitable for our estimation of the value of reddening.

The best value is determined by visual inspection of the plots for the best fit to the absorption feature at 2200 Å, representing the best agreement between standard theoretical (dashed line) values and observations (John Raymond, Private communication). The most suitable value of the reddening for II Peg is $E(B - V) = 0.10 \pm 0.02$ as shown in Fig.2.

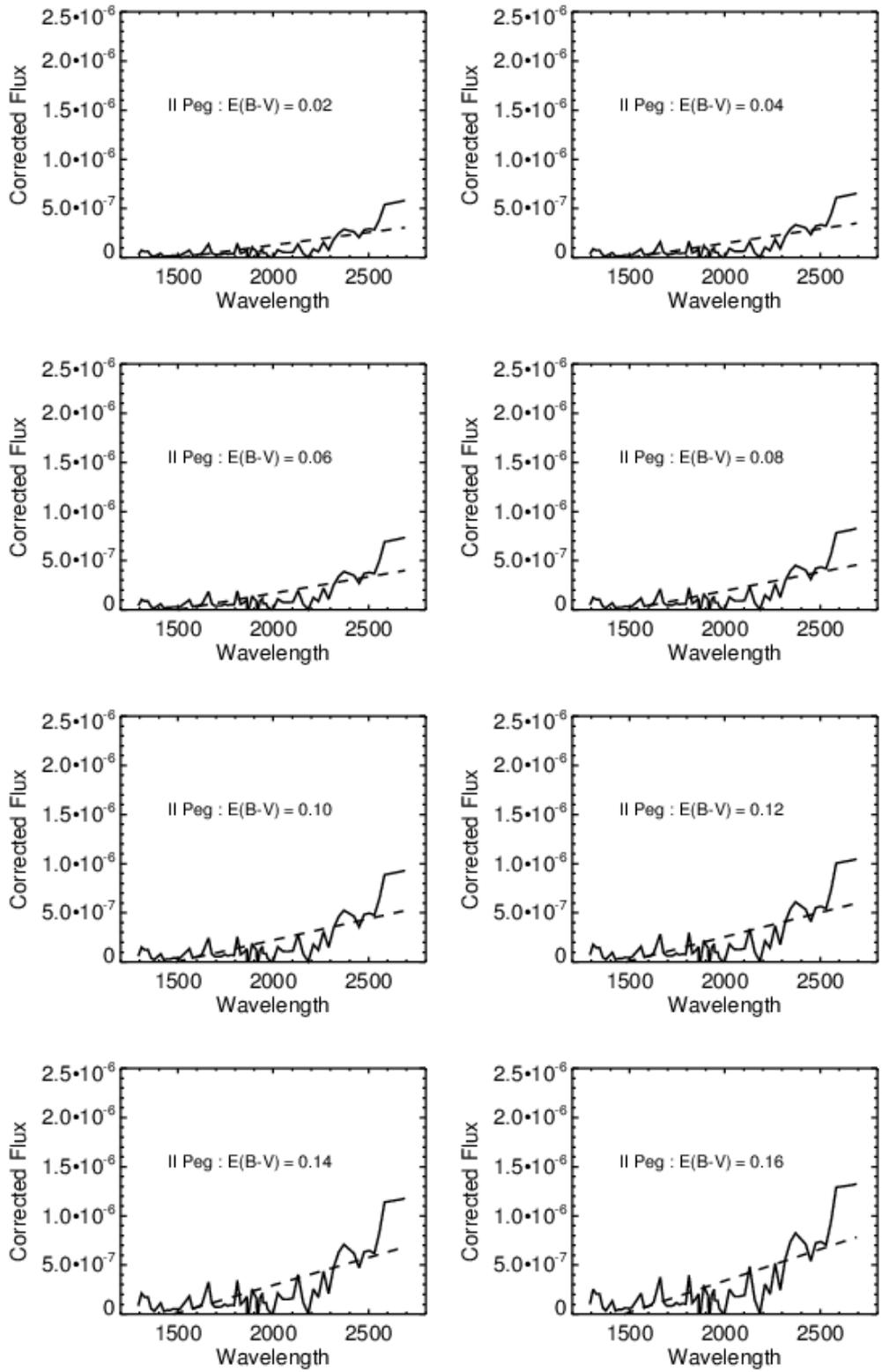

Fig.2. Reddening determination of II Peg binary system

## 3.2 Ultraviolet behaviour of spectral lines and its source

II Peg binary system show some ultraviolet emission lines observed with (IUE) Short Wavelength camera such as NV (1240Å), OI (1306Å),CII (1335Å), SIV (1400Å) and CIV (1550Å), He II (1640Å), CI (1657Å), SII (1808Å), Si III] (1892 Å).

The OI, Si IV, C IV and C I are resonance emission lines that are collisionally controlled by the physical conditions of plasma in the active chromosphere. The He II emission line at 1640 Å is a recombination line and the Si III emission line at 1808Å is collisionally excited line. These emission lines are originated in the chromosphere and transition region of the primary star, K2 IV previously discussed by (Andrews et al. 1988, Doyle et al. (1989), Sarro & Byrne (2000).

The similar behaviour of studied emission lines (OI, CII & CIV & He II, CI) suggest that they have the samesource of origination, the active chromosphere of the primaryobject. The line fluxes are treated as the integrated area included in the emission region above the continuum close to the wings of spectral line and calculated with the method of Gaussian profile fitting.

Figure 3 shows the behaviour of line fluxes with orbital phase for the spectral lines OI, CII, C IV, He II and CI. The fluxes of spectral lines correlate and vary with orbital phase with different values on short times of a few hours and long times of months and years for the period 1979 - 1993. The fluxes of spectral lines nearly vary by a factor between two and three. The fluxes increase at phases around (0.2, 0.45, 0.6, 0.7 and 0.9) and decrease at phases around (0.3, 0.4, 0.55, 0.85).

Tables (2, 3, 4, 5 & 6) show the line fluxes and flaring values of OI, C II, C IV, He II and C I emission lines for II Peg. We noticed thatthe occurrence of flaring activity of II Peg is not periodic as with UX Ari (Sanad, et al. 2021). The flux increase of studied emission lines reached about six and nine times the quiescent values and the activity of all mentioned emission lines is around phases (0.59 & 0.92)above the normal spectral behaviour as shown in Fig. 3.

Andrews et al. (1988) studied the spectral behaviour and flaring activity of II Peg for observations taken from IUE in 1983. They detected two flares and found that the increase in fluxes of flares reached about six times the quiescent value and the flares occurred at phases around 0.59and 0.86. They reported that the normal integrated line fluxes are approximately constant.

Sarro & Byrne (2000)studied the spectral behaviour of II Peg over the period 1979 – 1992 and concentrated mainly on the quiescent states of emission lines (CII, CIV, He II). They detected two flares of II Peg and compared them with the mean quiescent spectrum and found a dramatic increase in line fluxes.

The spectral behaviour of our studied spectral lines can be recognized as follows: The atmosphere of the primary objectK2 IV has strong magnetic fields leading to its non - radiative heating. The magnetic activity in stars is analogous to that observed with the Sun. Magnetic fields push hot particles upward leading to dark spots in the outer parts and reconnect and produce particles and magnetic energy as flares. Some of energy flares are believed to be a main source of heating the outer layers of the chromosphere. The existence of significant magnetic structures on the surface of the primary star have been confirmed by Zeeman-Doppler imaging (Carroll et al. 2009; Kochukhov et al. 2009). Though the relation between the magnetic fields and the dark spots is not yet evident, the spot activity is interpreted as a result of a magnetic dynamo working in the convective zone. The magnetic activity is variable with time (Messina 2008).

The fluxes of emission lines indicate a clear evidence of active regions in the chromosphere of II Peg. These active areas are close to the spot group with short term changes in the spot configuration. The increase in the fluxes of ultraviolet spectral lines indicating a hot area overlying a cool spot (Byrne et al. 1987, Rondono et al. 1986, Andrews et al. 1988, Kaluzney 1984)

The current ultraviolet observations with flaring activity can be explained by the model of two – ribbon flare in which the ribbons move quietly away as the flare increase. A prominence is often detected before the flare and vanishes at the beginning of the flare. The two ribbons form on either side of the prominence. As the ribbons move away loop structures are formed connecting one ribbon to the other known as post flare loops and indicating that there connection of magnetic field lines that accelerating particles to high energy and convert the magnetic energy into heat and kinetic energy has permitted the coronal magnetic field to relax into a lower energy state (Doyle et al. 1989, Mathioudakis1 et al. 2003).

A two - ribbon flare is connected with the existence of a filament in an unstable active area. The filament can be considered as a current wire with its associated magnetic field. The large inertia of the photospheric plasma prevents the magnetic field to penetrate through it, so the surface currents will be produced and as a result the magnetic field will be changed. The filament is adjusted by an equilibrium force. The motions of the photospheric layers result in an increase of filament current and the filament will move upward until at some critical height, the equilibrium is lost and a two - ribbon flare begins and the maximum storage of energy gained when the filament is situated between the two stars (Van Tend & Kuperus 1978, Van den Oord 1988, Doyle et al. 1989).

In summary: A flare requires a reconnection area and a reconnection area requires a current sheet and a current sheet requires an eruption of a rope of magnetic flux. The rope of magnetic flux representing a collection of magnetic field lines covering central axis is a significant structural factor in the process of reconnection eruption (Cheng et al. 2011). During the process of flare the electrons are accelerated and stream down along the loops of flare and heating the chromospheres and generating significant flare emission at the footpoints which form two ribbons. The ultraviolet ribbons with flare fluxes can be produced either by heating the energetic electrons accelerated in flares or by impact excitations by energetic electrons as non thermal excitation or by thermal conduction along loops (Forbes & Priest 1995, Fletcher 2002).

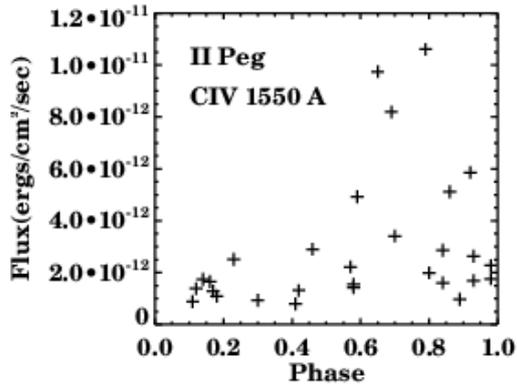
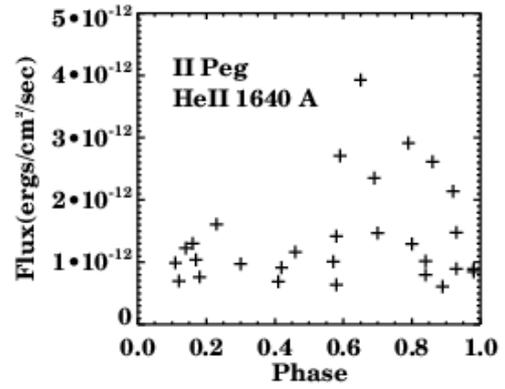
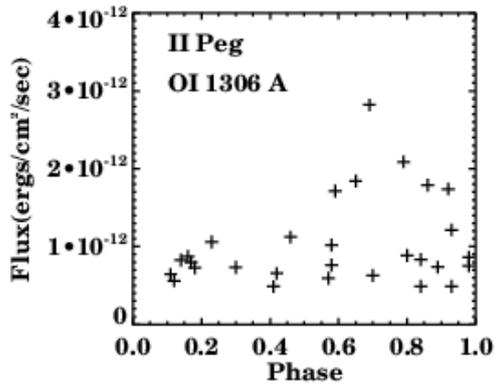
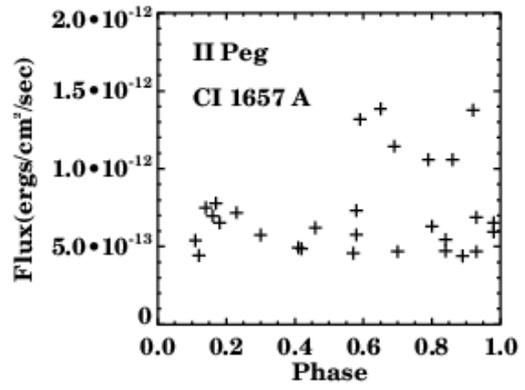
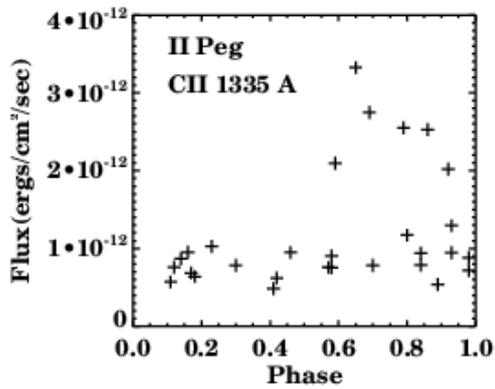

Fig. (3) Variation of the OI, CII, CIV, HeII and CI line fluxes with phase

## 3.3 Ultraviolet luminosity and the rate of mass loss

The ultraviolet luminosity in the selected lines is calculated by using the following equation

$$L_{UV} = 4\pi F d^2 \tag{3}$$

where F is the integrated flux value and d is the distance to the star 42 pc (Osten et al 2007) For II Peg, by using line fluxes of OI, C II, C IV, He II and C I, we found the ultraviolet luminosities for the five spectral lines in different states as listed in the Tables.

Table (2) Line fluxes and ultraviolet luminosities at different states of OI

| State | Flux (erg cm$^{-2}$s$^{-1}$) | L$_{uv}$ (erg s$^{-1}$) |
|---|---|---|
| Flare | 2.8 × 10$^{-12}$ | 5.96 × 10$^{29}$ |
| High | 1.2 × 10$^{-12}$ | 2.55 × 10$^{29}$ |
| Intermediate | 7.5 × 10$^{-13}$ | 1.59 × 10$^{29}$ |
| Low | 4.8 × 10$^{-13}$ | 1.02 × 10$^{29}$ |

Table (3) Line fluxes and ultraviolet luminosities at different states of CII

| State | Flux (erg cm$^{-2}$s$^{-1}$) | L$_{uv}$ (erg s$^{-1}$) |
|---|---|---|
| Flare | 3.3 × 10$^{-12}$ | 7.02 × 10$^{29}$ |
| High | 1.3 × 10$^{-12}$ | 2.76 × 10$^{29}$ |
| Intermediate | 8.7 × 10$^{-13}$ | 1.85 × 10$^{29}$ |
| Low | 4.8 × 10$^{-13}$ | 1.02 × 10$^{29}$ |

Table (4) Line fluxes and ultraviolet luminosities at different states of CIV

| State | Flux (erg cm$^{-2}$s$^{-1}$) | L$_{uv}$ (erg s$^{-1}$) |
|---|---|---|
| Flare | 1.1 × 10$^{-11}$ | 2.34 × 10$^{30}$ |
| High | 2.5 × 10$^{-12}$ | 5.32 × 10$^{29}$ |
| Intermediate | 1.3 × 10$^{-12}$ | 2.76 × 10$^{29}$ |
| Low | 7.9 × 10$^{-13}$ | 1.68 × 10$^{29}$ |

Table (5) Line flux and ultraviolet luminosities at different states of HeII

| State | Flux (erg cm$^{-2}$s$^{-1}$) | L$_{uv}$ (erg s$^{-1}$) |
|---|---|---|
| Flare | 3.9 × 10$^{-12}$ | 8.30 × 10$^{29}$ |
| High | 1.03 × 10$^{-12}$ | 2.19 × 10$^{29}$ |
| Intermediate | 8.4 × 10$^{-13}$ | 1.78 × 10$^{29}$ |
| Low | 6.3 × 10$^{-13}$ | 1.30 × 10$^{29}$ |

Table (6) values of line flux and ultraviolet luminosities at different states of CI

| State | Flux (erg cm$^{-2}$s$^{-1}$) | L$_{uv}$ (erg s$^{-1}$) |
|---|---|---|
| Flare | $1.3 \times 10^{-12}$ | $2.76 \times 10^{29}$ |
| High | $7.7 \times 10^{-13}$ | $1.63 \times 10^{29}$ |
| Intermediate | $5.9 \times 10^{-13}$ | $1.25 \times 10^{29}$ |
| Low | $4.4 \times 10^{-13}$ | $9.36 \times 10^{28}$ |

The rate of mass loss is calculated by using the following equation (Nieuwenhuijzen & de Jager)

$$\log(M^\bullet) = 14.02 + 1.24\log\left(\frac{L}{L_\odot}\right) + 0.16\log\left(\frac{M}{M_\odot}\right) + 0.81\log\left(\frac{R}{R_\odot}\right) M_\odot \text{yr}^{-1} \quad (4)$$

where the mass of the primary object M ~0.8M$_\odot$ (Osten et al 2007) and a radius of ~3R$_\odot$ (Berdyugina et al. 1998)

We found the rate of mass loss ~ $1 \times 10^{-8}$M$_\odot$yr$^{-1}$

## 4. Conclusions

The International Ultraviolet Explorer observations of II Peg binary system revealed similar behaviour of five emission lines (OI, CII, CIV, HeII & CI) confirming that they are originated in the same emitting area (the chromosphere and transition region of K star). The emitting source is distinguished by variations of the magnetic activity that is responsible for the observed phase-dependent variations in the line fluxes.

The flaring activity of II Peg is not periodic as with similar binary systems like UX Ari and the increment in fluxes of studied emission lines reached about six and nine times the quiescent values and the activity of all mentioned emission lines is around certain phases as reported in previous studies.

The estimated physical parameters (luminosity in ultraviolet, rate of mass loss) confirmed that the ultraviolet spectral lines are originated in the active chromosphere and transition region of the primary star.

The variations of spectral behaviour with orbital phase support the model of two – ribbon flare where the reconnection of magnetic field lines that accelerating particles to high energy and convert the magnetic energy into heat and kinetic energy has permitted the coronal magnetic field to relax into a lower energy state